\documentclass[letterpaper,aps,prc,superscriptaddress,nofootinbib,showpacs,floatfix]{revtex4-1}
\usepackage{hyperref}
\usepackage[utf8]{inputenc}
\usepackage{amsmath}
\usepackage{amsfonts}
\usepackage{amssymb}
\usepackage{amsthm}
\usepackage{bm}
\usepackage[english]{babel}
\usepackage{fontenc}
\usepackage{graphicx}
\usepackage[margin=1in]{geometry}
\usepackage{enumerate}
\usepackage{textcomp}
\usepackage[toc,page]{appendix}
\usepackage{slashed}
\usepackage{color}
\usepackage{xcolor}
\usepackage{setspace}
\graphicspath{ {figure/}}
\hypersetup{colorlinks=true, linkcolor=blue, citecolor=blue, urlcolor=blue}

{\large }

\begin{document}
\title{Unified Balance Functions}

\author{ Claude~Pruneau }
\email{claude.pruneau@wayne.edu}
\affiliation{Department of Physics and Astronomy, Wayne State University, Detroit, 48201, USA}
\author{ Victor~Gonzalez }
\email{victor.gonzalez@cern.ch}
\affiliation{Department of Physics and Astronomy, Wayne State University, Detroit, 48201, USA}
\author{ Brian~Hanley }
\email{bghanley@wayne.edu}
\affiliation{Department of Physics and Astronomy, Wayne State University, Detroit, 48201, USA}
\author{Ana Marin} 
\email{a.marin@gsi.de}
\affiliation{GSI Helmholtzzentrum f\"ur Schwerionenforschung, Research Division and ExtreMe Matter Institute EMMI, Darmstadt, Germany}
\author{ Sumit~Basu }
\email{sumit.basu@cern.ch}
\affiliation{Lund University, Department of Physics, Division of Particle Physics, Box 118, SE-221 00, Lund, Sweden}
\date{Sept 2022}

\begin{abstract}
The use of charge balance functions in heavy-ion collision studies  was initially proposed as a probe of delayed hadronization and two-stage quark production in these collisions. It later emerged that general balance functions can also serve as a probe of the diffusivity of light quarks as well as the evolution of 
the systems formed in heavy-ion collisions. In this work, we reexamine the formulation of general balance functions and consider how to best define and measure these correlation functions in terms of differences of  conditional densities of unlike-sign and like-sign particle pairs. We  define general balance functions in terms of associated particle functions  and show these obey a simple sum rule. We additionally proceed to distinguish between balance functions expressed as differences of conditional densities valid irrespective of experimental acceptance boundaries and bound balance functions that explicitly account for the limited acceptance of experiments. General balance functions are additionally extended to accommodate  strange, baryon, as well as charm and bottom quantum numbers based on the densities of these quantum numbers.
\end{abstract}

\maketitle

\section{Introduction}

Balance functions (BFs) were introduced  in the study of  heavy-ion collisions at RHIC as a tool to investigate the evolution of particle production with collision centrality~\cite{Bass:2000az,Jeon:2001ue}, and more specifically, to seek evidence of delayed hadronization and two-stage quark emission in these collisions. More recently, it was also shown that BFs may serve as a probe of the diffusivity of light quarks~\cite{Pratt:2019pnd,Pratt:2021xvg} as well as the chemical evolution of the hot matter formed in A--A collisions~\cite{Pratt:2015jsa,PhysRevC.99.044916}. The light quark diffusivity (LQD) was found to impact the shape and width of azimuthal projections of BFs: the larger the diffusivity is, the larger the BF become azimuthally ($\Delta\varphi$) as a result of light quark scatterings during the short lifetime of the dense QGP systems formed in heavy-ion collisions~\cite{Pratt:2019pnd,Pratt:2021xvg}. However, the shape of BFs is also influenced by a number of other phenomena, including the fraction of late (vs. early) quark production determined by the temperature of the system~\cite{Bass:2000az,Jeon:2001ue,Basu:2016ibk}, the presence of strong pressure gradients and the rapid transverse expansion of the QGP matter~\cite{Pruneau:2007ua,Bialas200431,Basu:2020jbk}, quantum statistic   effects (i.e., HBT)~\cite{Pratt:2022xbk}, as well as  feed down from resonance decays~\cite{Pratt:2015jsa}. In spite of these caveats,  measurements of general  balance functions do provide a new and  complementary approach towards the determination of viscous effects and the diffusivity of light quarks~\cite{ALICE:2021hjb}. While studies of flow performed on the basis azimuthal multi-particle correlations are driven, in large collision systems, by the collision geometry and somewhat hampered by non-flow effects, the estimation of the diffusivity and viscous effects with   balance functions is less dependent on knowledge of  the collision geometry and relies explicitly  on two-particle correlations and the impact of the medium  on these correlations. Conclusions reached with the two approaches  should thus yield mutually compatible values of these observables~\cite{ALICE:2021hjb,STAR:2015ryu,STAR:2003kbb}.

\begin{figure}[!ht]
	\centering
	\includegraphics[width=0.99\linewidth,trim={2mm 3mm 2mm 4mm},clip]
	{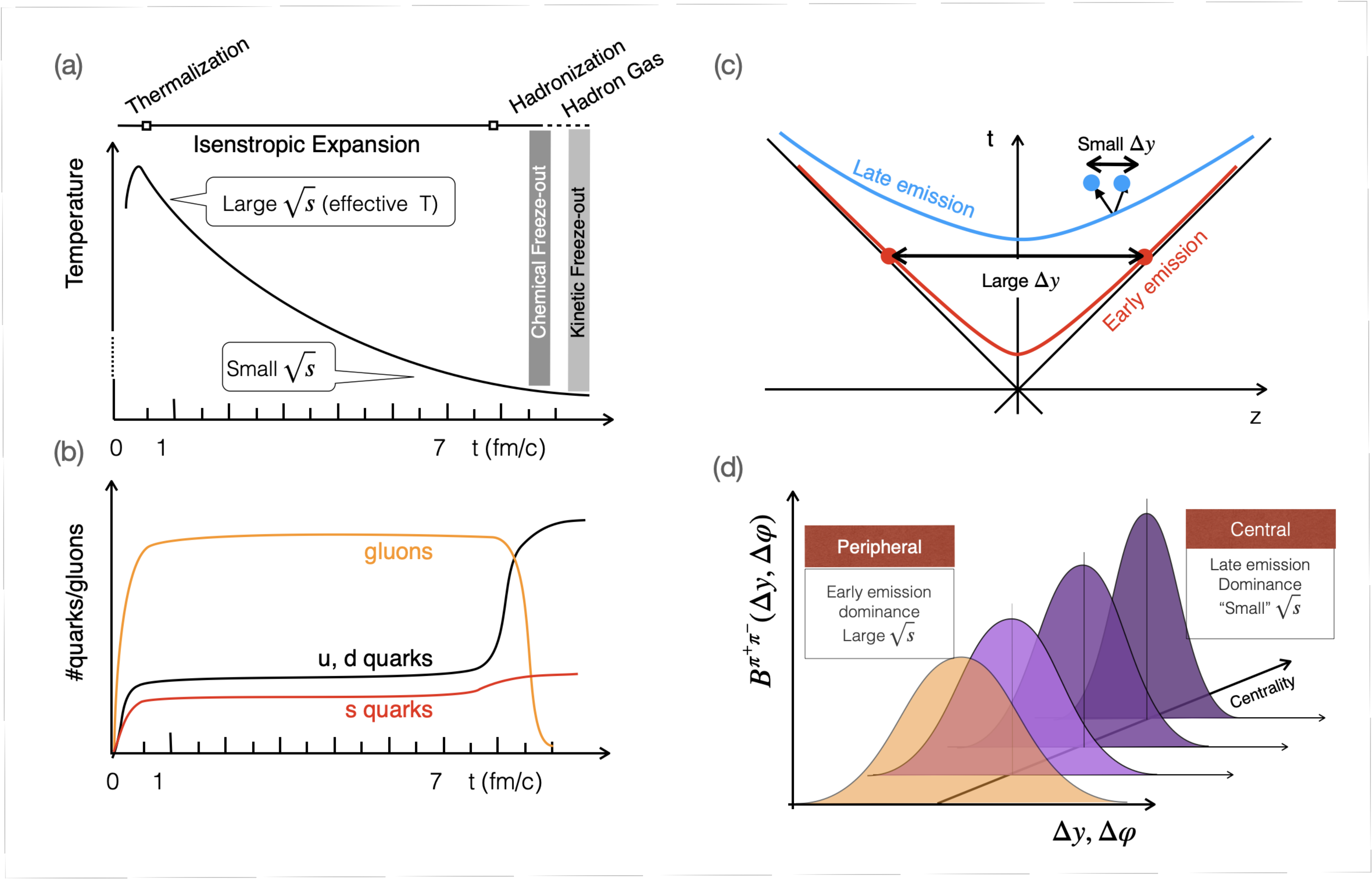}
\caption{(a) Schematic evolution of  A--A collision system (effective) temperature vs. time. The variable $\sqrt{s}$ here represents the average collision energy of quarks and gluons at at a given time. In equilibrated systems, it is determined by the temperature and decreases as the system expands; (b) schematic abundance of gluons and quarks vs. time; (c) Schematic representation of early and late quark production and its impact of the relative rapidity of particle pairs; (d) Expected evolution of balance functions, plotted as  functions of rapidity (pseudorapidity) and azimuth differences vs.  collision centrality. }
	\label{fig:Canonical} 
\end{figure}

Panels (a,b) of Fig.~\ref{fig:Canonical} schematically represent the time evolution of the  system temperature and the abundance of quarks and gluons commonly assumed to take place in collisions of heavy-ions featuring a substantial quark gluon plasma (QGP) component and an extended isentropic expansion stage~\cite{Bass:2000az,SCARDINA2013296,Fukushima_2016}. Strange (charm, bottom) quarks being heavier, their production shall preferentially occur at early times featuring the highest effective temperature whereas lighter up and down quarks can be abundantly produced at late stages of the collision (as well as early times) as the system hadronizes. The variable $\sqrt{s}$  represents the average effective collision energy of quarks and gluons at a given time during the collision. In locally thermalized system, $\sqrt{s}$ is determined by the effective temperature of the system~\cite{Gardim:2019xjs,Basu:2016rdk}. As the temperature decreases, so does $\sqrt{s}$ and the particles created by collisions accordingly  feature  smaller average longitudinal rapidity differences. Panel (c) schematically shows the relative  effects of early and late emission of $q\bar q$ pairs on the rapidity difference of hadrons they eventually  produce, whereas panel (d) qualitatively  illustrates the evolution of the shape of balance functions on the collision centrality as a result of changes in the early/late quark emission dominance and the narrowing effect engendered by radial flow. Not shown are  effects of scattering (diffusivity) of quarks and hadrons, which are expected to produce a broadening of the $\Delta \varphi$ width of balance functions~\cite{Pratt:2019pnd,Pratt:2021xvg}. In the absence of a QGP component or with a very short lived isentropic expansion stage, all particles would be produced at about the same time and average $\sqrt{s}$ and one would thus expect no substantial change of the balance function widths vs. collision centrality. In the other extreme, i.e., if the system is fully thermalized, memory of the quark  production time and mechanisms  is  lost thereby  resulting in very broad and featureless balance functions. 

Precise determination of the diffusivity of light quarks and other properties  shall  require one properly controls and corrects measurements of the shape and integral of BFs. Indeed, determining the diffusivity of light quarks and other properties of the QGP will require careful comparison of high precision measurements with detailed calculations of the evolution of nuclear matter and its impact on the shape and strength of BFs~\cite{Basu:2020ldt}. It should additionally be noted that many  heavy-ion  models currently in use in the field, particularly those  assuming  grand-canonical particle production or a hydrodynamic expansion phase followed by Cooper Fry particlization do not and cannot produce realistic balance functions. Further development and deployment of balance function measurements shall thus open the door to a better understanding of the microscopic plasma. 

One must also note that the notion of charge balance function can readily be extended to general (charge) balance functions involving identified particle species as well as baryon number and strangeness balance functions, explicitly discussed for the first time in this work. It is thus important for theoreticians and experimentalists to clearly define the correlation functions known as {\it general balance functions}  and agree on specific definitions and notations of the theoretical quantities being measured and their actual implementation in  measurements. It is  the purpose of this work to explore definition options and propose specific choices of formulations and notations of general balance functions as standards for use by the community.  In this work, our goal is to assert what constitutes, theoretically, the most  meaningful definition of general balance functions, and  experimentally, the best approach to measure them and their integrals. 

Notations for the different components involved in the elaboration of  BFs are defined in sec.~\ref{sec:definitions:densities}, whereas  the notion of general balance functions is introduced in  sec.~\ref{sec:definitions:ibf} based on integral quantities. The notion of general balance function is extended to   correlation  functions of  pairs of identified particle species  in sec.~\ref{sec:definitions:bf}. Charge conservation and the presence of net charge imply a BF sum rule discussed in sec.~\ref{sec:BFsumRules}. This naturally leads to extensions involving baryon and strangeness balance functions
in sec.~\ref{sec:baryonStrangenessBF}. Experimental considerations, involving, in particular, measurements of balance function in difference coordinates, e.g., $\Delta y$ and $\Delta\varphi$, are discussed in sec.~\ref{sec:acceptance}. We also briefly discuss, in sec.~\ref{sec:BalFctVsNudyn} the connection between balance functions and the $\nu_{\rm dyn}^{+-}$ observable~\cite{Pruneau:2002yf}. This work is summarized in sec.~\ref{sec:summary}.

\section{Notation and Definitions}
\label{sec:definitions:densities}

Herein, the identity of  particle species (e.g., $\pi^{+}$, $\rm K^{+}$, etc) is represented  with Greek letters $\alpha$, $\beta$, etc, and their respective anti-particles (e.g., $\pi^{-}$, $\rm K^{-}$, etc) with barred letters $\bar \alpha$, $\bar\beta$. 
 
Single and pair densities of species $\alpha$ and $\beta$ are denoted and defined according to

\begin{align}
\rho_1^{\alpha}(y_1,\varphi_1,p_{\rm {T,1}}) &\equiv  \frac{d^3N_1^{\alpha}}{dy_1d\varphi_1 dp_{\rm T,1}}, \\ 
\rho_2^{\alpha\beta}(y_1,\varphi_1,p_{\rm {T,1}},y_2,\varphi_2,p_{\rm {T,2}}) &\equiv \frac{d^6N_2^{\alpha\beta}}{dy_1d\varphi_1 dp_{\rm T,1}dy_2d\varphi_2 dp_{\rm T,2}} 
\end{align}

where $N_1^{\alpha}$ and $N_2^{\alpha\beta}$ respectively represent numbers of particles of species $\alpha$  and pairs of particles of species $\alpha$ and $\beta$. Variables $y_{1}$, $\varphi_1$, $p_{\rm T,1}$ and $y_{2}$, $\varphi_2$, $p_{\rm T,2}$  are the rapidity, azimuth, and transverse momentum  of particles of species $\alpha$ and $\beta$, respectively.  

The average number of particles of species $\alpha$, measured  per event, within an acceptance $\Omega$  is 

\begin{align} 
\label{eq:AvgDensityRho1}
\langle N_1^{\alpha}\rangle  &= \int_{\Omega}  \rho_1^{\alpha}(y,\varphi,p_{\rm T}) dy d\varphi dp_{\rm T} = V \bar{\rho}_1
\end{align}
where $V=\int_{\Omega}dy d\varphi dp_{\rm T}$ is the selected/accepted phase space volume and $\bar\rho_1^{\alpha}$ the  average density across this volume. Similarly, the average number of {\bf pairs} of particles of species $\alpha$ and $\beta$, measured within $\Omega$, is given by 

\begin{align}
	\label{eq:AvgDensityRho2}
\langle N_2^{\alpha\beta}\rangle &= \left\langle N_1^{\alpha}\left( N_1^{\beta} - \delta_{\alpha\beta} \right) \right\rangle = \int_{\Omega}dy_1 d\varphi_1 dp_{\rm T,1}\int_{\Omega} dy_2 d\varphi_2 dp_{\rm T,2} \hspace{0.05in}\rho_2^{\alpha\beta}(y_{1},\varphi_1,p_{\rm T,1},y_{2},\varphi_2,p_{\rm T,2}) 
\end{align}

In the following, if a particular variable, e.g., $p_{\rm T}$, is omitted from the expression of densities, it is assumed to be integrated across the fiducial acceptance of the detector. For instance,

\begin{align}
	\rho_1^{\alpha}(y_1,\varphi_1) &= \int_{\Omega_1} dp_{\rm T,1}  \rho_1^{\alpha}(y_1,\varphi_1,p_{\rm {T,1}}), \\ 
	\rho_2^{\alpha\beta}(y_1,\varphi_1,y_2,\varphi_2)  &= \int_{\Omega_1} dp_{\rm T,1}  \int_{\Omega_2}  dp_{\rm T,2}\, \rho_2^{\alpha\beta}(y_1,\varphi_1,p_{\rm {T,1}},y_2,\varphi_2,p_{\rm {T,2}}),
	\end{align}
where $\Omega$, with $i=1,2$, represent the $p_{\rm T}$ acceptance of particles of type $\alpha$ and $\beta$, respectively. For the sake of simplicity, the rapidity  acceptance of the measurement shall be assumed, herewith, to be the same for all particles species: $-y_0 \le y < y_0$. 

The averages $\langle N_1^{\alpha}\rangle$ and $\langle N_2^{\alpha\beta}\rangle$  correspond to first and second factorial moments and are hereafter denoted $f_1^{\alpha}$ and $f_2^{\alpha\beta}$, respectively~\cite{Pruneau:2017ypa}. It is also useful to consider first and second order  factorial cumulants, $F_1^{\alpha}$ and $F_2^{\alpha\beta}$, computed respectively as 

\begin{align}
	\label{eq:factorialCumulants1}
    F_1^{\alpha} &=  f_1^{\alpha}, \\ 
	\label{eq:factorialCumulants2}
    F_2^{\alpha\beta}&= f_2^{\alpha\beta} - f_1^{\alpha}f_1^{\beta},
\end{align}
as well as   normalized second order cumulants $R_2^{\alpha\beta}$  defined as

\begin{align}
	\label{eq:normalizedFactorialCumulants}
    R_2^{\alpha\beta} &= \frac{ F_2^{\alpha\beta}}{F_1^{\alpha}F_1^{\beta}} 
    = \frac{\langle N_2^{\alpha\beta}\rangle}{\langle N_1^{\alpha}\rangle\langle N_1^{\beta}\rangle} -1,
\end{align}
where it is implicitly assumed that all integral quantities are determined within the measurement acceptance $\Omega$.

\section{Integral Balance Functions}
\label{sec:definitions:ibf}

We first consider the definition of general balance functions (BF) based on integral quantities. Rather than defining BFs based on combinations of $+-$, $-+$, $++$, and $--$
particle pairs as in Ref.~\cite{Pratt:2002BFLH}, we ``split" the definition to consider $+-$ and  $-+$ pairs relative to $--$ and $++$ pairs, respectively. The two definitions should evidently be equivalent for symmetric collision systems. Experimentally, however, instrumental artifacts may induce artificial  differences between $+-$ and  $-+$ pairs and it is thus of interest to explicitly verify that the two definitions yield the same value thereby enabling validation of experimental calibrations and correction methods~\cite{Pruneau:2019baa}.

Hereafter, we shall use the notation $I^{\alpha\bar\beta}$ for integral balance functions, which correspond, as we shall see, to integrals across the measurement acceptance of  differential balance functions denoted $B^{\alpha\bar\beta}(y_1,y_2)$  defined in sec.~\ref{sec:definitions:bf}.

Let us tentatively define general charge (integral) balance functions  according to
\begin{align}
\label{eq:B2alphaBarBetaInt}
I^{\alpha\bar\beta} &= \frac{\langle N_2^{\alpha\bar\beta}\rangle}{\langle N_1^{\bar\beta}\rangle}
- \frac{\langle N_2^{\bar\alpha\bar\beta}\rangle}{\langle N_1^{\bar\beta}\rangle} \\ 
\label{eq:B2BarAlphaBetaInt}
I^{\bar\alpha\beta} &= \frac{\langle N_2^{\bar\alpha\beta}\rangle}{\langle N_1^{\beta}\rangle}
- \frac{\langle N_2^{\alpha\beta}\rangle}{\langle N_1^{\beta}\rangle}
\end{align}
which should give us a measure of how many particles of type $\alpha$($\bar{\alpha}$) balance each ``trigger" particle $\bar{\beta}$($\beta$).
One straightforwardly verifies these expressions converge to unity
for $\alpha=+,\bar\beta=-$ and $\bar\alpha=-,\beta=+$, i.e., 
\begin{align}
    I^{+-} \rightarrow 1, \\
    I^{-+} \rightarrow 1,
\end{align}
in the ideal limit of a $4\pi$ detection system with full $p_{\rm T}$ coverage
and  collisions  involving a vanishing net charge $Q$, e.g., $p\bar p$ collisions. Indeed, for $\alpha,\beta=+$ and $\bar\alpha,\bar\beta=-$, by virtue of charge conservation, the creation of a particle of type $\alpha=+$ must be accompanied by the production  of a particle of type $\bar\alpha=-$. If the number of such pair creations (i.e., number of sources) is $N_s$  in a given event, then the {\bf total} number of singles and pairs are 
\begin{align}
\label{NvsNs}
    N_1^{+}& = N_s, \\
    N_1^{-}& = N_s, \\ 
    N_2^{+-}& = N_s^2, \\ 
    N_2^{-+}& = N_s^2, \\ 
    N_2^{++}& = N_s(N_s-1), \\ 
    N_2^{--}& = N_s(N_s-1). 
\end{align}
The expressions~(\ref{eq:B2alphaBarBetaInt},\ref{eq:B2BarAlphaBetaInt}) computed over the full $4\pi$ acceptance (all rapidities and transverse momenta) thus indeed converge to unity
\begin{align}
I^{-+}(4\pi)= I^{+-}(4\pi) &=  
 \frac{\langle N_s^2\rangle}{ \langle N_s\rangle}
- \frac{\langle N_s^2 -N_s\rangle}{\langle N_s\rangle} = 1.
\end{align}
However, the above definitions, Eqs.~(\ref{eq:B2alphaBarBetaInt},\ref{eq:B2BarAlphaBetaInt}), do not account for the presence, ab initio, of a non-vanishing net-charge $Q$. For instance, $pp$ collisions feature $Q=2$ ab initio and given the electric charge is a conserved quantity, the event-wise single and pair yields shall be
\begin{align}
\label{NvsNsWithQ}
    N_1^{+}& = N_s + Q, \\
    N_1^{-}& = N_s, \\ 
    N_2^{+-}& = (N_s+Q)N_s, \\ 
    N_2^{-+}& = N_s(N_s+Q), \\ 
    N_2^{++}& = (N_s+Q) (N_s+Q-1), \\ 
    N_2^{--}& = N_s(N_s-1),
\end{align}
in each event.
The definitions (\ref{eq:B2alphaBarBetaInt},\ref{eq:B2BarAlphaBetaInt}) thus yield
\begin{align}
\label{eq:B2+-WithQ}
I^{+-}(4\pi) &= \frac{\langle (N_s+Q)N_s \rangle}{ \langle N_s\rangle}
- \frac{\langle N_s (N_s+-1) \rangle}{\langle N_s\rangle} = 1+ Q, \\ 
\label{eq:B2-+WithQ}
I^{-+}(4\pi) &=  
  \frac{\langle (N_s+Q)N_s \rangle}{ \langle N_s + Q\rangle}
- \frac{\langle (N_s+Q) (N_s+Q-1) \rangle}{\langle N_s + Q\rangle} = 1 -Q,
\end{align}
where the notation $(4\pi)$ indicates the integral is computed in full angular and $p_{\rm T}$ acceptance. The presence of the terms $Q$ and $-Q$ in the above two equations results from charge conservation and the initial net charge $Q$. 
It implies, for instance, that the integral of the  p$\rm \bar{p}$ BF measured in pp collisions could amount to  +3 or -1 depending on trigger species. Similarly, Pb--Pb collisions could yield BF integral amounting to 1+(82+82)= 165 or 1-(82+82)=-163. Evidently, the impact of the non vanishing net charge should  be less important in the central rapidity region  when the beam rapidity is very large (e.g, LHC energies) but could be significant at low RHIC/BES energies  that involve  beam rapidities of  order 4 or smaller. The presence of non-vanishing net charge may then confuse the interpretation of balance functions and their integrals. It is then convenient to eliminate this dependence and modify the definition of integral balance functions, Eqs.~(\ref{eq:B2alphaBarBetaInt},\ref{eq:B2BarAlphaBetaInt}), according to
\begin{align}
\label{eq:B2alphaBarBetaIntNoQ}
I^{\alpha\bar\beta} &\equiv  \frac{\langle N_2^{\alpha\bar\beta}\rangle}{\langle N_1^{\bar\beta}\rangle}
- \frac{\langle N_2^{\bar\alpha\bar\beta}\rangle}{\langle N_1^{\bar\beta}\rangle} - \left( \langle N_1^{\alpha} \rangle - \langle N_1^{\bar\alpha} \rangle\right)\\ 
\label{eq:B2BarAlphaBetaIntNoQ}
I^{\bar\alpha\beta} &\equiv \frac{\langle N_2^{\bar\alpha\beta}\rangle}{\langle N_1^{\beta}\rangle}
- \frac{\langle N_2^{\alpha\beta}\rangle}{\langle N_1^{\beta}\rangle} + \left( \langle N_1^{\alpha} \rangle - \langle N_1^{\bar\alpha} \rangle\right)
\end{align}
which shall, by construction, yield   $I^{+-}\rightarrow 1$,  $I^{-+}\rightarrow 1$ in the full $4\pi$ and $p_{\rm T}$ acceptance limit.

Experimentally, a full acceptance is not achievable, and one might be limited to, e.g., $-y_0 \le y < y_0$ and $p_{\rm T,\min} \le p_{\rm T} < p_{\rm T,\max}$, with full azimuthal acceptance\footnote{The discussion is formulated in terms of particle rapidities but readily also applies to pseudorapidities}. One straightforwardly verifies that the  balance functions (\ref{eq:B2alphaBarBetaIntNoQ}, \ref{eq:B2BarAlphaBetaIntNoQ})  computed 
 within such limited  acceptance $\Omega$  shall be smaller than unity. 
It is also useful to note that  
 $I^{\alpha\bar\beta}$ and $I^{\bar\alpha\beta}$  
can also be expressed in terms  of integral cumulants and normalized integral cumulants (sometimes called reduced cumulants) according to 
\begin{align}
\label{eq:B2alphaBarBetaIntVsR2}
I^{\alpha\bar\beta}  &= F_1^{\alpha}  R_2^{\alpha\bar\beta}  - F_1^{\bar\alpha}  R_2^{\bar\alpha\bar\beta}, \\ 
\label{eq:B2BarAlphaBetaIntVsR2}
I^{\bar\alpha\beta} &= F_1^{\bar\alpha} R_2^{\bar\alpha\beta} - F_1^{\alpha} R_2^{\alpha\beta}.
\end{align}
Hereafter, we shall denote the arithmetic average of  $I^{\alpha\bar\beta}$ and $I^{\bar\alpha\beta}$ as $I^{\alpha\beta,s}$
\begin{align}
\label{eq:B2alphaBarBetaIntVsR2V2}
I^{\alpha\beta,s}  &= \frac{1}{2}\left( I^{\alpha\bar\beta} + I^{\bar\alpha\bar\beta}\right).
\end{align}
It is evidently clear, by virtue of Eqs.~(\ref{eq:B2+-WithQ},\ref{eq:B2-+WithQ}), that $I^{\alpha\beta,s}$ shall converge to unity in the full $4\pi$ and $p_{\rm T}$ acceptance limit, irrespective of the net charge $Q$ of the system.

\section{Differential Balance Functions}
\label{sec:definitions:bf}

With the definitions (\ref{eq:B2alphaBarBetaIntNoQ}, \ref{eq:B2BarAlphaBetaIntNoQ}) in hand, we consider the formulation of balance function based on conditional densities $\rho_2^{\alpha|\beta}(y_1|y_2)$~\cite{Bass:2000az} computed according to  
\begin{equation}
    \rho_2^{\alpha|\beta}(y_1|y_2) = \frac{ \rho_2^{\alpha\beta}(y_1,y_2)  }{ \rho_1^{\beta}(y_2)}.
\end{equation}
By construction,  $\rho_2^{\alpha|\beta}(y_1|y_2)$ amounts  to the density of a species $\alpha$ at $y_1$ given a particle of species $\beta$ is detected at $y_2$\footnote{Hereafter, for simplicity and without loss of generality,  densities and balance functions are written as functions of rapidity only. }. To simplify the discussion, we first neglect the net charge $Q$ and write differential balance function according to 
\begin{align}
\label{eq:B2abCondition1}
   B^{\alpha|\bar\beta}(y_1|y_2) &=  \rho_2^{\alpha|\bar\beta}(y_1|y_2) - \rho_2^{\bar\alpha|\bar\beta}(y_1|y_2) \\
\label{eq:B2abCondition2}
   &  = \frac{ \rho_2^{\alpha\bar\beta}(y_1,y_2)  }{ \rho_1^{\bar\beta}(y_2)}
   - \frac{ \rho_2^{\bar\alpha\bar\beta}(y_1,y_2)  }{ \rho_1^{\bar\beta}(y_2)}
\end{align} 
which is to be considered a function of $y_1$ {\bf only} since $y_2$ is  ``given". Particle $\bar\beta$, found at $y_2$, is considered the ``trigger" particle whereas particles $\alpha$ and $\bar \alpha$, detected at $y_1$, are called ``associated" particles. While it is intuitively tempting to think of the function $B^{\alpha|\bar\beta}(y_1|y_2)$ as  the density of particles of type $\alpha$ at $y_1$ {\bf given} a particle of type $\bar\beta$ is found at $y_2$, one must acknowledge that $B^{\alpha|\bar\beta}(y_1|y_2)$ can in fact be negative across some fraction of the domain $y_1$ and thus does not amount, strictly speaking,  to a particle density.

Considering once again the basic case of an inclusive charge balance function, e.g., $\alpha=\beta=+$, one writes 
\begin{align}
\label{eq:B2+-Density}
   B^{+|-}(y_1|y_2)  &  = \frac{ \rho_2^{+-}(y_1,y_2)  }{ \rho_1^{-}(y_2)}
   - \frac{ \rho_2^{--}(y_1,y_2)  }{ \rho_1^{-}(y_2)}
\end{align}
Since $y_2$ is given, one can then proceed to integrate $B^{+|-}(y_1|y_2)$ over $y_1$. The production of a negatively charged particle must be accompanied by the production of a  positively charged one somewhere in phase space. The integral of the balance function $B^{+|-}(y_1|y_2)$, denoted 
\begin{equation}\label{I2}
    I^{+|-}(y_2|\Omega) \equiv \int_{\Omega} dy_1 B^{+|-}(y_1|y_2),
\end{equation}
thus  converges to unity, by construction,  in the $4\pi$ and full $p_{\rm T}$ acceptance limit:
\begin{equation}
\label{eq:limit4Pi}
    \lim_{\Omega\rightarrow 4\pi} I^{+|-}(y_2|\Omega) \rightarrow 1.
\end{equation}
 Evidently, in that limit, $I^{+|-}(y_2|\Omega)$ has the same value for all   $y_2$. However, for a given $y_2$ and a finite acceptance $\Omega: -y_0 \le y < y_0$, the integral $I^{+|-}(y_2)$ shall in general depend on $y_0$. Consider that if the given value is $y_2=0$ and the acceptance of the measurement is symmetric $-y_0 \le y < y_0$, it is obviously easier to  ``catch" the balancing  partner than if the given position is $y_2=y_0$. Indeed, in that case, balancing partners can only be found on ``one side" whereas for $y_2=0$, balancing partners can be found on two sides.  
One thus concludes  the integral  $I^{+|-}(y_2)$ is a function of $y_2$ which depends on the shape and width of $B^{+|-}(y_1|y_2)$. It thus makes sense to   consider the {\bf average} of $I^{+|-}(y_2)$   across the acceptance $\Omega$ of the measurement:\footnote{For simplicity, we  assume a symmetric acceptance $-y_0 \le y < y_0$}
\begin{equation}
    \bar{I}^{+-}(y_0) \equiv \int_{-y_0}^{y_0} dy_2 P_1^-(y_2) I^{+|-}(y_2),
\end{equation}
where $P_1^-(y_2)$ represents the probability of finding the first particle at $y_2$.
Clearly, this probability is 
\begin{equation}
    P_1^{-}(y_2) = \frac{1}{\langle N_1^-\rangle(y_0)}  \rho_1^{-}(y_2),
\end{equation}
which, by construction, satisfies $\int_{-y_0}^{y_0} dy_2 P_1^-(y_2) = 1$.
The average sought for is thus
\begin{equation}
	\label{eq:I2PMvsy1y2}
    \bar{I}^{+-} = \frac{1}{\langle N_1^-\rangle} \int_{-y_0}^{y_0} dy_2 
    \int_{-y_0}^{y_0} dy_1 \left[ 
    \rho_2^{+-}(y_1,y_2) -  \rho_2^{--}(y_1,y_2)
    \right]
        = \frac{1}{\langle N_1^-\rangle}\left[
    \langle N_2^{+-}\rangle - \langle N_2^{--}\rangle \right],
\end{equation}
which is identical in form to Eq.~(\ref{eq:B2alphaBarBetaInt}) for $\alpha =+$, $\bar\beta=-$ when $Q=0$. It thus becomes natural to define the BF as a joint function of $y_1$ and $y_2$ according to 
\begin{equation}
	\label{eq:B2vsy1y2}
    B^{+-}(y_1,y_2|y_0) = \frac{1}{\langle N_1^-\rangle}\left[ \rho_2^{+-}(y_1,y_2) - \rho_2^{--}(y_1,y_2) \right],
\end{equation}
the integral of which yields Eq.~(\ref{eq:B2alphaBarBetaInt}). 
The same reasoning, repeated for $B^{-+}(y_1,y_2|y_0)$, yields
\begin{equation}
	\label{eq:B2vsy1y2Bis}
    B^{-+}(y_1,y_2|y_0) = \frac{1}{\langle N_1^+\rangle}\left[ \rho_2^{-+}(y_1,y_2) - \rho_2^{++}(y_1,y_2) \right],
\end{equation}
The expression (\ref{eq:B2vsy1y2}) was derived based on Eq.~(\ref{eq:B2+-Density}), and thus neglects the presence of a non-vanishing net charge $Q$. For $Q\ne 0$, integration of 
$B^{+|-}(y_1|y_2)$ over the full phase space $\Omega\rightarrow 4\pi$ shall then  yield $1+Q$   rather than $1$. However note that by definition, integration of the difference $\rho_1^+(y) - \rho_1^-(y)$   yields the net charge $Q$. To obtain a balance function definition that integrates to 1, even in the presence of $Q\ne 0$, it thus suffices to subtract this difference  from Eq.~(\ref{eq:B2+-Density}). Repeating the same reasoning for $B^{-|+}(y_1|y_2)$,
one thus proceed to define charge balance functions according to
\begin{align}
\label{eq:B2+-NoQ}
   B^{+|-}(y_1|y_2) &= \frac{ \rho_2^{+-}(y_1,y_2)  }{ \rho_1^{-}(y_2)}
   - \frac{ \rho_2^{--}(y_1,y_2)  }{ \rho_1^{-}(y_2)}
   - \left[ \rho_1^+(y_1) - \rho_1^-(y_1) \right]\\
\label{eq:B2-+NoQ}
   B^{-|+}(y_1|y_2) &=  \frac{ \rho_2^{-+}(y_1,y_2)  }{ \rho_1^{+}(y_2)}
   - \frac{ \rho_2^{++}(y_1,y_2)  }{ \rho_1^{+}(y_2)}
   + \left[ \rho_1^+(y_1) - \rho_1^-(y_1) \right].
\end{align}
These expressions are defined at a given value of $y_2$ and must thus be averaged over the acceptance $\Omega$ to yield a BF defined for all values of $y_1$ and $y_2$. Proceeding as above, one takes the averages of $B^{+|-}(y_1|y_2)$ and $B^{-|+}(y_1|y_2)$ weighed 
by the probabilities $P_1^{\alpha}(y_2)=\rho_1^{\alpha}(y_2)/\langle N_1^{\alpha}\rangle$  of finding a particle of type $\alpha$ at $y_2$, for $\alpha = -, +$, respectively. This yields ``bound" balance functions
\begin{align}
\label{eq:B2+-y1y2NoQ}
   B^{+-}(y_1,y_2|y_0) &= 
   \frac{1}{\langle N_1^-\rangle}\left[ \rho_2^{+-}(y_1,y_2) - \rho_2^{--}(y_1,y_2) 
   -  \rho_1^+(y_1)\rho_1^-(y_2) + \rho_1^-(y_1)\rho_1^-(y_2) \right]\\
\label{eq:B2-+y1y2NoQ}
   B^{-+}(y_1,y_2|y_0) &=    \frac{1}{\langle N_1^+\rangle}\left[ \rho_2^{-+}(y_1,y_2) - \rho_2^{++}(y_1,y_2) 
   -  \rho_1^-(y_1)\rho_1^+(y_2) + \rho_1^+(y_1)\rho_1^+(y_2) \right].
\end{align}
By construction, these integrate to unity in the $4\pi$ (full $p_{\rm T}$ coverage) limit even in the presence of a non-vanishing net charge, i.e., $Q\ne 0$. Noting the presence of terms of the form $\rho_2^{\alpha\beta}-\rho_1^{\alpha}\rho_1^{\beta}$, it is convenient to write the above expressions 
as
\begin{align}
\label{eq:B2+-y1y2NoQ-C}
   B^{+-}(y_1,y_2|y_0) &= 
   \frac{1}{\langle N_1^-\rangle}\left[ C_2^{+-}(y_1,y_2) - C_2^{--}(y_1,y_2) \right]\\
\label{eq:B2-+y1y2NoQ-C}
   B^{-+}(y_1,y_2|y_0) &=    \frac{1}{\langle N_1^+\rangle}\left[ C_2^{-+}(y_1,y_2) - C_2^{++}(y_1,y_2) \right],
\end{align}
where we introduced the differential correlation functions $C_2^{\alpha\beta}$
defined according to 
\begin{equation}
\label{eq:C2}
    C_2^{\alpha\beta}(y_1,y_2) = \rho_2^{\alpha\beta}(y_1,y_2) - \rho_1^{\alpha}(y_1)\rho_1^{\beta}(y_2).
\end{equation}


The expressions Eqs.~(\ref{eq:B2+-NoQ},\ref{eq:B2-+NoQ}) were defined for charge balance functions but their structure does not limit their applicability to inclusive measurements and we show in sec.~\ref{sec:BFsumRules} they obey a simple sum rule which also conserves charges and accounts for the net charge of the system. It is thus appropriate to introduce general balance functions according to 
\begin{align}
\label{eq:B2GenAlphaBarBeta}
      B^{\alpha|\bar\beta}(y_1|y_2) &= A_2^{\alpha|\bar\beta}(y_1|y_2)- 
      A_2^{\bar\alpha|\bar\beta}(y_1|y_2), \\ 
\label{eq:B2GenBarAlphaBeta}
      B^{\bar\alpha|\beta}(y_1|y_2) &= A_2^{\bar\alpha|\beta}(y_1|y_2)- 
      A_2^{\alpha|\beta}(y_1|y_2),
  \end{align}
  where we introduced single ``associated" particle functions
  according to 
\begin{align}
\label{eq:A2}
      A_2^{\alpha|\beta}(y_1|y_2) &= \frac{C_2^{\alpha\beta}(y_1|y_2)}{\rho_1^{\beta}(y_2)}
=  \frac{\rho_2^{\alpha\beta}(y_1,y_2)}{\rho_1^{\beta}(y_2)}   
 - \rho_1^{\alpha}(y_1).
\end{align}
It should first be noted that $A_2^{\alpha|\bar\beta}(y_1|y_2)$ and
$B^{\alpha|\bar\beta}(y_1|y_2)$ are single particle and single variable functions, given the rapidity $y_2$ is considered given and thus not a free variable in the context of the definitions in Eqs.~(\ref{eq:B2GenAlphaBarBeta},\ref{eq:B2GenBarAlphaBeta},\ref{eq:A2}). 
Additionally, by construction, and in the absence of correlations, the density $\rho_2^{\alpha\beta}(y_1,y_2)$ shall factorize according to
\begin{equation}
    \rho_2^{\alpha\beta}(y_1,y_2) = \rho_1^{\alpha}(y_1)\rho_1^{\beta}(y_2).
\end{equation}
The associated particle function $A_2^{\alpha|\beta}(y_1|y_2)$ shall then vanish, by definition, for independent particle emission (i.e., no correlations). However, in the presence of correlations, the pair density $\rho_2^{\alpha\beta}(y_1,y_2)$
may be larger or smaller than $\rho_1^{\alpha}(y_1)\rho_1^{\beta}(y_2)$ over some 
kinematic domain of $y_1$ and $y_2$. The function $A_2^{\alpha|\beta}(y_1|y_2)$ may then be positive, negative, or null across some portions of the acceptance.
It is similarly straightforward to observe that the balance functions may also be negative or null across some portions of the acceptance. As such, neither $A_2^{\alpha|\beta}(y_1|y_2)$ nor $B^{\alpha|\beta}(y_1|y_2)$  can  be considered single particle densities. It should be additionally noted that the shape and strength of $A_2^{\alpha|\beta}(y_1|y_2)$ and thus $B^{\alpha|\beta}(y_1|y_2)$
may depend strongly  on $y_2$. For instance, at rapidity $y_2$ near the beam rapidity $y_{\rm B}$, one expects the particle production to be largely dominated by the fragmentation of the beam components whereas at central rapidity ($y\approx 0$
in a collider mode), particle production is determined by large $\sqrt{s}$ processes. The widths and shapes of BFs are thus indeed expected to vary appreciably with the selected rapidity $y_2$.

Experimentally, measurements of (general) balance functions are restricted to finite ranges of rapidity, transverse momentum, as well as, in some cases, azimuth. The general balance functions Eqs.~(\ref{eq:B2GenAlphaBarBeta}, \ref{eq:B2GenBarAlphaBeta}) must then be ``averaged" for the  position of the trigger particle: $y_2$, $p_{\rm T,2}$, and $\varphi_2$. Repeating the steps leading to Eqs.~(\ref{eq:B2+-y1y2NoQ-C}, \ref{eq:B2-+y1y2NoQ-C}), one gets the bound general balance functions defined according to
\begin{align}
\label{eq:B2alphaBarBetay1y2NoQ-C}
   B^{\alpha\bar\beta}(y_1,y_2|\Omega) &= 
   \frac{1}{\langle N_1^{\bar\beta}\rangle}\left[ C_2^{\alpha\bar\beta}(y_1,y_2) - C_2^{\bar\alpha\bar\beta}(y_1,y_2) \right]\\
\label{eq:B2BarAlphaBetay1y2NoQ-C}
   B^{\bar\alpha\beta}(y_1,y_2|\Omega) &=    \frac{1}{\langle N_1^{\beta}\rangle}\left[ C_2^{\bar\alpha\beta}(y_1,y_2) - C_2^{\alpha\beta}(y_1,y_2) \right],
\end{align}
which are applicable to same, $\alpha=\beta$, or mixed, $\alpha\ne \beta$,  particle species, each carrying a single unit of charge.

It is worth mentioning that Eqs.~(\ref{eq:B2alphaBarBetay1y2NoQ-C}, \ref{eq:B2BarAlphaBetay1y2NoQ-C}) are not applicable to physical systems involving multiply charged particles, i.e., when particles of type $\alpha$, $\beta$ may be multi-charge species, such as $\Delta^{++}$ or $^4$He, and so on. In such cases, one must replace the single and pair particle densities, $\rho_1^\alpha$ and $\rho_2^{\alpha\beta}$,  by single and pair electric charge densities defined according to 
\begin{align}
    \rho_{\rm e 1}^{\alpha} &= n_{\rm e}^{\alpha} \rho_{1}^{\alpha} \\
    \rho_{\rm e 2}^{\alpha\beta} &= n_{\rm e}^{\alpha} n_{\rm e}^{\beta} \rho_{2}^{\alpha\beta}
\end{align}
where $n_{\rm e}^{\alpha}$ and $n_{\rm e}^{\beta}$ represent the number of elementary charges of species $\alpha$ and $\beta$, respectively. Correspondingly, for cases where $\alpha$ and $\beta$ correspond to specific particle species,  Eqs.~(\ref{eq:B2+-y1y2NoQ}, {\ref{eq:B2-+y1y2NoQ}}) transform  to
\begin{align}
\label{eq:B2+-y1y2NoQ_mc}
   B^{\alpha\bar\beta}(y_1,y_2|\Omega) &= 
   \frac{n_{\rm e}^{\alpha}}{\langle N_1^{\bar\beta}\rangle}\left[ 
   \rho_2^{\alpha\bar\beta}(y_1,y_2) -
   \rho_2^{\bar\alpha\bar\beta}(y_1,y_2) 
   -  \rho_1^\alpha(y_1)\rho_1^{\bar\beta}(y_2) + \rho_1^{\bar\alpha}(y_1)\rho_1^{\bar\beta}(y_2) \right]\\
\label{eq:B2-+y1y2NoQ_mc}
   B^{\bar\alpha\beta}(y_1,y_2|\Omega) &=    \frac{n_{\rm e}^{\alpha}}{\langle N_1^{\beta}\rangle}\left[ \rho_2^{\bar\alpha\beta}(y_1,y_2) - \rho_2^{\alpha\beta}(y_1,y_2) 
   -  \rho_1^{\bar\alpha}(y_1)\rho_1^{\beta}(y_2) + \rho_1^{\alpha}(y_1)\rho_1^{\beta}(y_2) \right].
\end{align}
When particles of type $\alpha$, $\beta$ include different species with different number of elementary charges, e.g., $+$,
$++$, and  $+++$, the single and pair electric charge densities shall then be defined according to
\begin{align}
\label{eq:sumOfCharges1}
    \rho_{\rm e 1}^{\alpha} &= \sum_{\gamma} n_{\rm e}^{\gamma} \rho_{1}^{\gamma} \\
\label{eq:sumOfCharges2}
    \rho_{\rm e 2}^{\alpha\beta} &= \sum_{\gamma}\sum_{\mu} n_{\rm e}^{\gamma} n_{\rm e}^{\mu} \rho_{2}^{\gamma\mu}
\end{align}
where $\gamma$($\mu$) refers to particles of type $\alpha$($\beta)$ with an elementary charge of $n_{\rm e}^{\gamma}$($n_{\rm e}^{\mu}$) units.
Eqs.~(\ref{eq:B2+-y1y2NoQ}, {\ref{eq:B2-+y1y2NoQ}}) then transform  to
\begin{align}
\label{eq:B2+-y1y2NoQ_mcms}
   B^{\alpha\bar\beta}(y_1,y_2|\Omega) &= 
   \frac{1}{\sum_{\bar\mu} n_{\rm e}^{\bar\mu}\langle N_1^{\bar\mu}\rangle}\left[ 
   \rho_{\rm e 2}^{\alpha\bar\beta}(y_1,y_2) -
   \rho_{\rm e 2}^{\bar\alpha\bar\beta}(y_1,y_2) 
   -  \rho_{\rm e 1}^\alpha(y_1)\rho_{\rm e 1}^{\bar\beta}(y_2) + \rho_{\rm e 1}^{\bar\alpha}(y_1)\rho_{\rm e 1}^{\bar\beta}(y_2) \right]\\
\label{eq:B2-+y1y2NoQ_mcms}
   B^{\bar\alpha\beta}(y_1,y_2|\Omega) &= 
   \frac{1}{\sum_{\mu} n_{\rm e}^{\mu}\langle N_1^{\mu}\rangle}\left[ \rho_{\rm e 2}^{\bar\alpha\beta}(y_1,y_2) - \rho_{\rm e 2}^{\alpha\beta}(y_1,y_2) 
   -  \rho_{\rm e 1}^{\bar\alpha}(y_1)\rho_{\rm e 1}^{\beta}(y_2) + \rho_{\rm e 1}^{\alpha}(y_1)\rho_{\rm e 1}^{\beta}(y_2) \right].
\end{align}

\section{Balance Functions Sum Rules}
\label{sec:BFsumRules}

Can the notion of balance function duly  apply to mixed species of particle? Do the definitions, Eqs.~(\ref{eq:B2GenAlphaBarBeta},\ref{eq:B2GenBarAlphaBeta}) properly account for charge conservation and the charge of the system?  

One expects, for instance, that the emission of a negative pion, $\pi^{-}$, shall be balanced  by the production of a positive ($+$ve) particle. Such a $+$ve particle could of course be a $\pi^{+}$, but it does not have to be. Indeed balancing the charge of the $\pi^{-}$ can be accomplished, in part, via the production of a $\rm K^{+}$, a proton ($\rm p$), or some other positively charged particle. In general, particles with masses larger than the mass of the proton tend to decay into either $\pi^{+}$,  $\rm K^{+}$, $\rm p$, or some positive weakly decaying particle. Such weak decays may lead to the production 
of $\pi^{+}$,  $\rm K^{+}$, $\rm p$, or   positrons $\rm e^+$. The balance function   $B^{+|\pi^-}$, which loosely speaking  corresponds to the ``probability" of finding a charge balancing partner to the $\pi^-$  shall thus amount to the sum of  balance functions $B^{\alpha|\pi^-}$ that involve particle of type $\alpha$ charge balancing the $\pi^-$:
\begin{equation}
\label{eq:sumRule1}
    B^{+|\pi^-}(y_1|y_2) = \sum_{\alpha} B^{\alpha|\pi^-}(y_1|y_2), 
\end{equation}
where the sum on $\alpha$ spans all particle species that potentially balance the production of a $\pi^-$. Evidently, if the sum rule applies  to the ``theoretical" balance functions $B^{+|\pi^-}(y_1|y_2)$, it shall apply also, by virtue of its derivation, to the bound (experimental) functions $B^{+\pi^-}(y_1,y_2)$.

We show, in the next paragraph, that the sum rule, Eq.~(\ref{eq:sumRule1}), does apply, by construction, to any other types of positive (negative) particle species $\beta$ ($\bar\beta$):
\begin{align}
\label{eq:sumRuleGeneric+}
    B^{+|\bar\beta}(y_1|y_2) &= \sum_{\alpha} B^{\alpha|\bar\beta}(y_1|y_2), \\ 
\label{eq:sumRuleGeneric-}
    B^{-|\beta}(y_1|y_2) &= \sum_{\alpha} B^{\bar\alpha|\beta}(y_1|y_2). 
\end{align}
Such balance functions sum rules have already been considered in the context of net proton number fluctuations for a system with vanishing net charge~\cite{Pruneau:2019BNC} but are here extended to include the presence of non-vanishing net charge in a collision system. 

In the remainder of this section, which can be omitted in a first reading, we show that the definitions~(\ref{eq:B2+-NoQ},\ref{eq:B2-+NoQ}), and 
the charge conservation limit, Eq.~(\ref{eq:limit4Pi}),
imply 
\begin{align}
\label{eq:limit+4PiAlphaBeta}
\lim_{\Omega\rightarrow 4\pi} I^{+|\bar\beta}(y_2) &= \int dy_1  B^{+|\bar\beta}(y_1|y_2) \rightarrow 1, \\ 
\label{eq:limit-4PiAlphaBeta}
\lim_{\Omega\rightarrow 4\pi} I^{-|\beta}(y_2) &= \int dy_1  B^{-|\beta}(y_1|y_2) \rightarrow 1. 
\end{align}
The definitions (\ref{eq:B2+-NoQ},\ref{eq:B2-+NoQ}) thus not only account for charge conservation but also properly handle the presence of net charge. The derivation is carried out for $B^{+|\bar\beta}(y_1|y_2)$ but evidently trivially applies to $B^{-|\beta}(y_1|y_2)$.

The derivation of the sum rule, Eq.~(\ref{eq:sumRuleGeneric+}), based on the definition, Eq.~(\ref{eq:B2+-NoQ}),  is accomplished by partitioning the single and pair densities according to 
\begin{align} 
\label{eq:decomp1}
\rho_1^{+} &= \sum_{\alpha} \rho_1^{\alpha}, \\ 
\rho_1^{-} &= \sum_{\alpha} \rho_1^{\bar\alpha}, \\ 
\rho_2^{+-} &= \sum_{\beta}\sum_{\alpha}  \rho_2^{\alpha\bar\beta}, \\ 
\label{eq:decomp2}
\rho_2^{--} &= \sum_{\beta}\sum_{\alpha} \rho_2^{\bar\alpha\bar\beta},
\end{align}
where sums span all species or anti-species as appropriate. and arguments $y_1$ and $y_2$ were omitted to simplify the notation. The integral $I^{+-}$, computed in the full acceptance limit, may then be written
\begin{align} \label{eq:def}
1 &=\int dy_1 B^{+-}(y_1|y_2) \\ 
&= \int dy_1 \left\{  \frac{(\rho_2^{+-} -\rho_2^{--})}{\rho_1^-} - \rho_1^+ + \rho_1^- \right\}.
\end{align} 
Inserting the decompositions, Eqs.~(\ref{eq:decomp1}-\ref{eq:decomp2}),
one gets 
\begin{align} 
1 &=\int dy_1 \frac{1}{\sum_{\beta} \rho_1^{\bar\beta}}
\left[ 
\sum_{\beta}\sum_{\alpha} \left(  \rho_2^{\alpha\bar\beta} -\rho_2^{\bar\alpha\bar\beta} \right)
\right]
- \sum_{\alpha} \left( \rho_1^{\alpha} - \rho_1^{\bar\alpha} \right) 
\end{align} 
Multiplying the first term within brackets  by $1=\rho_1^{\bar\beta}/\rho_1^{\bar\beta}$ and the second term by $1=\sum_{\beta} \rho_1^{\bar\beta}/\sum_{\beta} \rho_1^{\bar\beta}$, and rearranging the sums, one obtains
\begin{align} 
1 &=\int dy_1 \frac{1}{\sum_{\beta} \rho_1^{\bar\beta}}
\left\{ 
\sum_{\beta}\sum_{\alpha} \left[  \frac{\rho_1^{\bar\beta}\left(\rho_2^{\alpha\bar\beta} -\rho_2^{\bar\alpha\bar\beta}\right)}{\rho_1^{\bar\beta}} - \rho_1^{\bar\beta} \left( \rho_1^{\alpha} - \rho_1^{\bar\alpha} \right) \right] 
\right\} 
\end{align} 
Extracting $\rho_1^{\bar\beta}$ from the sum $\sum_{\alpha}$, one gets 
\begin{align} 
1 &=\int dy_1 \frac{1}{\sum_{\beta} \rho_1^{\bar\beta}}
\left\{ 
\sum_{\beta} \rho_1^{\bar\beta} \sum_{\alpha} \left[  \frac{\left(\rho_2^{\alpha\bar\beta} -\rho_2^{\bar\alpha\bar\beta}\right)}{\rho_1^{\bar\beta}} -  \left( \rho_1^{\alpha} - \rho_1^{\bar\alpha} \right) \right] 
\right\}, 
\end{align}
in which one  identifies the expression within the square brackets as $B^{\alpha\bar\beta}(y_1|y_2)$. Swapping the order of the sum and the integral, one finally gets 

\begin{align} 
1 &= 
\sum_{\beta}  \frac{\rho_1^{\bar\beta}}{\sum_{\beta'} \rho_1^{\bar\beta'}}\int dy_1\left\{  \sum_{\alpha} B^{\alpha\bar\beta} 
\right\}.  
\end{align}
which is true, in general, i.e., for any number of partitions $\alpha$ and $\beta$ if and only if
\begin{align} 
\label{eq:SumRuleAlphaBeta}
1 &=\int dy_1 \sum_{\alpha} B^{\alpha\bar\beta} =    \sum_{\alpha} \int dy_1 B^{\alpha\bar\beta}
\end{align}

The sum  $\sum_{\alpha} B^{\alpha\bar\beta}$, which spans all $+$ve species, thus indeed integrates to 1 and the sum-rule is proven. Experimentally, in a limited acceptance,  this  sum still corresponds to $B^{+|\bar\beta}(y_1|y_2)$ but the functions does not integrate to unity: the components $B^{\alpha\bar\beta}$ partition the sum $B^{+|\bar\beta}(y_1|y_2)$ and their contribution to this sum is a function of the size of the acceptance and the specific processes that lead to the join production of species $\alpha$ and $\bar\beta$.

\section{Baryon Number and Strangeness Balance Functions}
\label{sec:baryonStrangenessBF}

The notion of balance function is readily extended to baryon, strangeness, and charm balance functions. One must however account for the baryon number, strangeness number, or charm  carried by the particles. 

The baryon number of elementary hadrons is nominally confined to a minimal set of values $(-1, 0, 1)$ and hadrons with a null baryon number (i.e., mesons) are to be ignored in the computation of baryon balance functions. The computation of baryon balance functions shall then  nominally be restricted to hadrons with baryons number with  $B=1$ and anti-baryons with  $B=-1$.  However, it is well known that baryons produced in heavy-ion collisions may bind to form light nuclei (e.g., $^2$H, $^3$He, $^4$He and their respective anti-nuclei). Such  $B=A$ and $B=-A$ baryons and anti-baryons should thus nominally be included in the computation and measurements of baryon balance functions. However, the production of light-nuclei and anti-light-nuclei  at central rapidities  is a relatively rare occurrence. Nuclei  and anti-nuclei may then likely be neglected, at least in first approximation, in the computation of baryon number balance functions. 

Nominally,  sum rules of the form~(\ref{eq:sumRuleGeneric+}, \ref{eq:sumRuleGeneric-}) should apply  to baryon balance functions. Unfortunately, the detection of neutrons remains a significant challenge at collider energy. Contributions of the form $B^{\rm n|\bar\beta}$, where $\bar\beta$ represent a specific anti-baryon (e.g., anti-proton), shall thus be hard to assess. However,  partial balance functions  $B^{\rm p|\bar p}$, $B^{\rm n|\bar p}$, $B^{\Lambda|\rm \bar p}$, $B^{\Sigma|\rm \bar p}$, $B^{\Xi|\rm \bar p}$, and $B^{\Omega|\rm \bar p}$ should nearly exhaust balancing contributions to the production of  $\bar p$. The balance function sum rule, Eq.~(\ref{eq:sumRuleGeneric+}), shall then enable estimation of  $B^{\rm n|\bar p}$, which, in turn could be used to estimate cumulants of the neutron fluctuations~\cite{Braun-Munzinger:2019yxj}. 

The situation with strangeness balance functions is readily more complicated. First, one notes that multiply strange baryons, $s>1$, and anti-strange baryons, $s<-1$, may be produced in elementary particle or nucleus--nucleus  collisions. Accounting for the produced strangeness (or anti-strangeness) must then be based on  strangeness densities rather than number densities. Assuming the labels $\alpha$ and $\beta$ identify specific (unique) species, we define  single and pair strangeness densities  according to
\begin{align}
    \rho_{s1}^{\alpha} &= n_s^{\alpha} \rho_1^{\alpha}, \\ 
    \rho_{s2}^{\alpha\beta} &= n_s^{\alpha} n_s^{\beta} \rho_2^{\alpha\beta},
\end{align}
in which $n_s^{\alpha}$ and $n_s^{\beta}$ are the number of strange quarks (anti-quarks) in particles of type $\alpha$ and $\beta$, respectively. If the definitions of the labels $\alpha$ and $\beta$ each span several particle species, then
one must sum across these species as in Eq.~(\ref{eq:sumOfCharges1},\ref{eq:sumOfCharges2}) defined for electric charges to obtain single and pair densities.

Strangeness (unbound) balance functions can then be nominally computed as
\begin{align}
\label{eq:B2BandSGenAlphaBarBeta}
      B^{\alpha|\bar\beta}(y_1|y_2) &= \tilde{A}_2^{\alpha|\bar\beta}(y_1|y_2)- 
      \tilde{A}_2^{\bar\alpha|\bar\beta}(y_1|y_2), \\ 
\label{eq:B2BandSGenBarAlphaBeta}
      B^{\bar\alpha|\beta}(y_1|y_2) &= \tilde{A}_2^{\bar\alpha|\beta}(y_1|y_2)- 
      \tilde{A}_2^{\alpha|\beta}(y_1|y_2),
  \end{align}
  where we introduced {\bf strange}  ``associated" particle functions
  according to 
\begin{align}
\label{eq:BandSA2}
      \tilde{A}_2^{\alpha|\beta}(y_1|y_2) &=  \frac{\rho_{s2}^{\alpha|\beta}(y_1,y_2)}{\rho_{s1}^{\beta}(y_2)}   
 - \rho_{s1}^{\alpha}(y_1).
\end{align}
The second and more fundamental  difficulty arises from the kaon sector. Nominally, particle production yields charged kaons, $\rm K^{\pm}$, as well as neutral kaons, $\rm K^0$, and anti-kaons, $\rm \bar K^0$. The $\rm K^0$ and $\rm \bar K^0$ are however known to readily mix and yield weak eigenstates $\rm K^0_s$ and $\rm K^0_l$. The strangeness number of 
$\rm K^0_s$ and $\rm K^0_l$ is undefined (e.g., it is neither positive nor negative).  It is thus not possible to include the contributions of $\rm K^0$ and $\rm \bar K^0$ in balance functions to account for the production of strange and anti-strange quarks. Strange BFs shall thus be forever blind to the production of these two particles, which experimentally  materialize  as either $\rm K^0_s$ or $\rm K^0_l$. Measurements of strange balance functions in heavy-ion collisions  remain  nonetheless of great interest given the production of $s$ or $\bar s$ quarks is generally thought to feature a time evolution distinct of that of lighter quarks~\cite{Bass:2000az}. Quantitative comparisons of strange and charge balance functions may then enable better understanding and modeling of the collision dynamics and the properties of the QGP formed in A--A collisions.

Clearly, the notion of balance function can also be  applied to charmness or bottomness. Recent measurements have shown that measurements of correlation functions of charmed hadrons are in fact possible but it remains to be established whether such observations can be formulated as genuine charm balance functions~\cite{Basu:2021dzv,Vogt:2018oje,LHCb:2012aiv,Adolfsson:2020dhm}.

The existence of a gluon dominated phase at very early time of the  evolution of A--A collisions could provide significant insights and help distinguish the light and heavy quark evolution dynamics. Light quarks are more likely to be produced late in  collisions. The light hadrons they form are thus accordingly  less sensitive to early-time dynamics. By contrast, the production of  heavy quarks (strange, charm, bottom) requires higher $\sqrt{s}$ elementary collisions and is thus likely limited to early times. One expects that charm and bottom quarks being the heaviest, their production should be limited to very early times. Balance functions of open charm (bottom) particles should then reflect early time production and possibly heavy quark scattering within the QGP. However, 
given the mass of charm and bottom quarks are considerably heavier than those  of strange, up, and down quarks, they should be subjected  to smaller diffusivity effects~\cite{Pratt:2019pnd}. The balance function of charm might be then truly representative of early time collisions and one might expect a gradation of sensitivity to early times, that of charm and bottom being the largest, followed by strangeness, and  much less sensitivity from the lighter u and d quarks. 


\section{Balance Functions and Normalized Correlation Functions}
\label{sec:definitions:normalized}

Rather than conducting measurements of balance functions (and their integral) in terms 
of densities $\rho_2^{\alpha\beta}(y_1,y_2)$, it is also of interest to consider measurements based
on normalized differential two-particle cumulants defined according to

\begin{align}
 R_2^{\alpha\beta}(y_1,y_2) &\equiv\frac{ C_2^{\alpha\beta}(y_1,y_2)}{\rho_1^{\alpha}(y_1)\rho_1^{\beta}(y_2)} 
= \frac{ \rho_2^{\alpha\beta}(y_1,y_2)}{\rho_1^{\alpha}(y_1)\rho_1^{\beta}(y_2)} -1,
 \end{align}
 where the functions $C_2^{\alpha\beta}(y_1,y_2)$ are defined by Eq.~(\ref{eq:C2}).
 Analyses in terms of such normalized cumulants are of particular interest, experimentally, because this observable is robust against particle losses (efficiency) and thus, nominally, reduces the need for complicated efficiency correction procedures. 
The (unbound) balance functions, Eqs.~(\ref{eq:B2+-NoQ},\ref{eq:B2-+NoQ}),
may then be written
 \begin{align} 
 \label{eq:B2alphaBarBetaVsR2}
 B^{\alpha|\bar\beta}(y_1|y_2) &= \rho_1^{\alpha}(y_1) R_2^{\alpha\bar\beta}(y_1|y_2)
 - \rho_1^{\bar\alpha}(y_1) R_2^{\bar\alpha\bar\beta}(y_1|y_2),\\
 \label{eq:B2BarAlphaBetaVsR2}
 B^{\bar\alpha|\beta}(y_1|y_2) &= \rho_1^{\bar\alpha}(y_1) R_2^{\bar\alpha\beta}(y_1|y_2)
 - \rho_1^{\alpha}(y_1) R_2^{\alpha\beta}(y_1|y_2),
\end{align}
in which the normalized correlation functions $R_2$  are written with arguments of the form  $(y_1|y_2)$ to emphasize they are functions of $y_1$ given a value $y_2$.  However, given a specific acceptance $\Omega$, one can operationally define symmetric balance functions, i.e., function of two parameters $y_1$ and $y_2$ by averaging the integral of $B^{\alpha|\bar\beta}(y_1|y_2)$ and $B^{\bar\alpha|\beta}(y_1|y_2)$
across the acceptance of $y_2$. This is achieved  by averaging the integrals across the acceptance by weighing them with the probability to measure specific values of $y_2$. Proceeding as in sec.~\ref{sec:definitions:bf}, one then obtains bounded balance functions of the form 
\begin{align} 
 \label{eq:B2alphaBarBetaVsR2Sym}
 B^{\alpha\bar\beta}(y_1,y_2) &= 
 \frac{1}{\langle N_1^{\bar\beta}\rangle} 
 \left[
 \rho_1^{\alpha}(y_1)\rho_1^{\bar\beta}(y_2) R_2^{\alpha\bar\beta}(y_1,y_2) 
 - 
 \rho_1^{\bar\alpha}(y_1)\rho_1^{\bar\beta}(y_2) R_2^{\bar\alpha\bar\beta}(y_1,y_2) 
 \right], \\ 
  \label{eq:B2BarAlphaBetaVsR2Sym}
 B^{\bar\alpha\beta}(y_1,y_2) &=
 \frac{1}{\langle N_1^{\beta}\rangle} 
 \left[
 \rho_1^{\bar\alpha}(y_1)\rho_1^{\beta}(y_2) R_2^{\bar\alpha\beta}(y_1,y_2)
 - 
 \rho_1^{\alpha}(y_1)\rho_1^{\beta}(y_2) R_2^{\alpha\beta}(y_1,y_2) 
 \right],
\end{align}
in which the   $R_2$ are now written with arguments  of the form$(y_1,y_2)$ to indicate they
they are indeed functions of two parameters.

\section{Balance Functions vs. Invariant Momentum}
\label{sec:invariantMomentum}

\begin{figure}[!ht]
	\centering
	\includegraphics[width=0.9\linewidth,trim={2mm 3mm 2mm 4mm},clip]
	{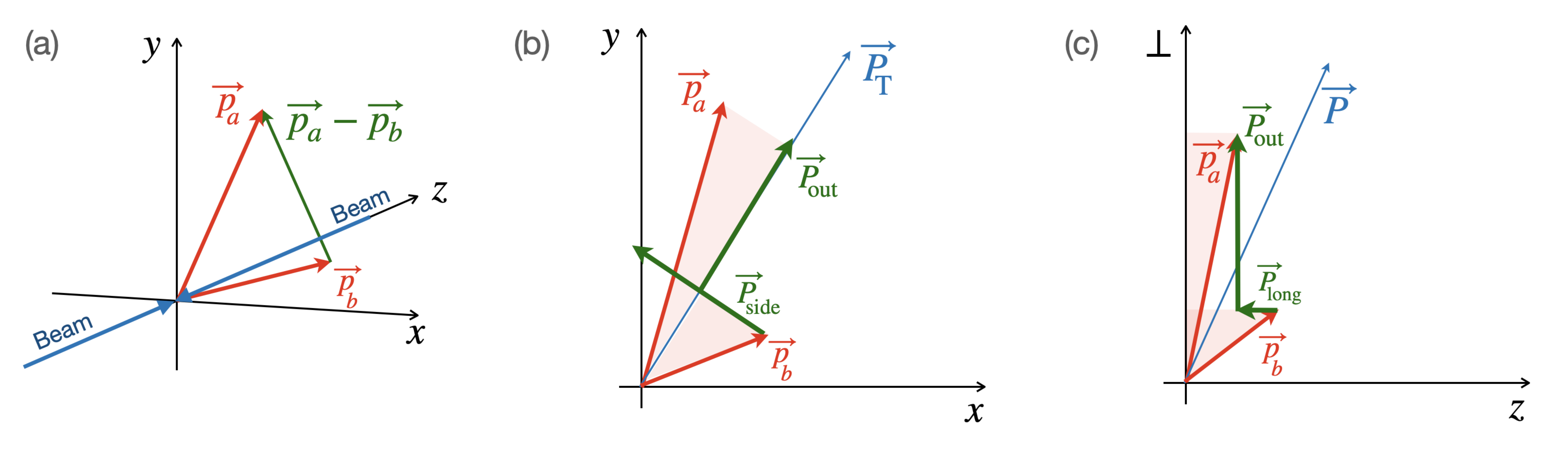}
\caption{ Schematic diagram of the pair differences $P_{\rm long}, P_{\rm out}, P_{\rm side}$ defined based on the particle momenta $\vec p_{a}$ and $\vec p_{b}$ with respect to the beam-axis and the total pair momentum $\vec P$ introduced   in the text. }
	\label{fig:Plongoutside} 
\end{figure}

The particle pair separation in momentum space is nominally determined by the energy of the process that produces a particular correlated pair. However, transport processes such as longitudinal and radial flow may affect the separation measured  in term of angular separation, e.g., azimuth angle pair separation, $\Delta \varphi$. The shape and strength of balance functions thus measured are influenced by both production and transport processes. In order to reduce this  causal  ambiguity, it may then be advantageous to carry out the BF measurements in terms of a relative  momentum invariants, $P_{\rm inv}$,  which is primarily determined by production processes and less affected by transport
phenomena such as  radial or longitudinal collective flow. To this end, Pratt et al. proposed BF measurements shall be carried in terms of particle pairs relative 4-momentum computed  in the reference frame of the two-particle center of mass according to~\cite{Pratt:2003inv}
\begin{align}
    \label{eq:qalphaPinvDef}
    q^{\mu}=(p_{a}^{\mu}-p_{b}^{\mu})-P^{\mu}\frac{P\cdot(p_{a}-p_{b})}{P^2}
    =(p_{a}^{\mu}-p_{b}^{\mu})-P^{\mu}\frac{m_a^2-m_b^2}{s},
\end{align}
in which $\mu=0,x,y,z$, $P$ is the total 4-momentum of the two particles  $P^{\mu}=p_{a}^{\mu}+p_{b}^{\mu}$, and the invariant $\sqrt{s}=\sqrt{(p_{a}+p_{b})^2}$ represents the center-of-mass (COM) energy of the pair. The square of the invariant momentum difference of the particles computed in the pair COM  frame is 
\begin{align}
    \label{eq:Pinvdef}
    P_{\rm inv}^2=-q^2=-(p_{a}-p_{b})^2+\frac{(m_a^2-m_b^2)^2}{P^2}.
\end{align}
Denoting  the two-particle transverse momentum,  $P_{\rm T}=\sqrt{P_{x}^2+P_{y}^2}$, it is convenient, as suggested by Pratt et al.~\cite{Pratt:2003inv}, to define three projections of the relative momentum according to  
\begin{align}
    \label{eq:Plongdef}
    P_{\rm long}&=\frac{1}{\sqrt{s+P_{\rm T}^{2}}}(P_{0}q_{z}-P_{z}q_{0}), \\
    \label{eq:Psidedef}
    P_{\rm side}&=\frac{P_{x}q_{y}-P_{y}q_{y}}{P_{\rm T}}, \\
    \label{eq:Poutdef}
    P_{\rm out}&=\sqrt{\frac{s}{s+P_{\rm T}^{2}}}\frac{P_{x}q_{x}+P_{y}q_{y}}{P_{\rm T}},
\end{align}
and such that 
\begin{align}
    \label{eq:Pinvlongsideout}
    P_{\rm inv}^2&=P_{\rm long}^2+P_{\rm side}^2+P_{\rm out}^2.
\end{align}
As illustrated in Fig.~\ref{fig:Plongoutside}, $\vec P_{\rm long}$ is the pair momentum difference along the beam axis (longitudinal separation), $\vec P_{\rm out}$ is along the two-particle transverse momentum $\vec P_{\rm T}$ (outwards separation), and $\vec P_{\rm side}$ points in the sidewards direction, i.e., in a direction perpendicular to both $P_{\rm long}$  and $P_{\rm out}$. 
The pair density in terms of $P_{\rm long}, P_{\rm out}, P_{\rm side}$ is
\begin{align}
    \label{PInvPairDensity}
    \rho_2^{\alpha\beta}(P_{\rm long}, P_{\rm out}, P_{\rm side})=\frac{{\rm d}^3N_2^{\alpha\beta}}{{\rm d}P_{\rm long}\,{\rm d}P_{\rm out}\,{\rm d}P_{\rm side}}.
\end{align}
Following a similar reasoning as that leading to Eq.~(\ref{eq:B2alphaBarBetay1y2NoQ-C}, \ref{eq:B2BarAlphaBetay1y2NoQ-C}), general balance functions may be written
\begin{align}
    \label{eq:InvBF+-}
    B^{\alpha\bar\beta}(P_{\rm long},P_{\rm out},P_{\rm side}|\Omega) &= \frac{1}{\langle N_1^{\bar\beta}\rangle}\left[ C_2^{\alpha\bar\beta}(P_{\rm long},P_{\rm out},P_{\rm side}) - C_2^{\bar\alpha\bar\beta}(P_{\rm long},P_{\rm out},P_{\rm side}) \right], \\ 
    \label{eq:InvBF-+}
    B^{\bar\alpha\beta}(P_{\rm long},P_{\rm out},P_{\rm side}|\Omega) &= \frac{1}{\langle N_1^{\beta}\rangle}\left[ C_2^{\bar\alpha\beta}(P_{\rm long},P_{\rm out},P_{\rm side}) - C_2^{\alpha\beta}(P_{\rm long},P_{\rm out},P_{\rm side}) \right],
\end{align}
in which 
\begin{align}
    \label{eq:C2Inv}
    C_2^{\alpha\beta}(P_{\rm long},P_{\rm out},P_{\rm side})= \rho_2^{\alpha\beta}(P_{\rm long},P_{\rm out},P_{\rm side}) - [\rho_1^{\alpha}\rho_1^{\beta}](P_{\rm long},P_{\rm out},P_{\rm side}).
\end{align}
The determination of BFs based on Eqs.~(\ref{eq:InvBF+-},\ref{eq:InvBF-+}) requires that measured pair yields $N_2^{\alpha\beta}(P_{\rm long},P_{\rm out},P_{\rm side}|\Omega)$ be fully corrected for efficiency losses to obtain densities $\rho_2^{\alpha\beta}(P_{\rm long},P_{\rm out},P_{\rm side})$ and correlation functions $C_2^{\alpha\beta}(P_{\rm long},P_{\rm out},P_{\rm side})$. Alternatively, experimentally,  it may be  preferable to compute the BFs in terms of normalized cumulants
\begin{align}
    \label{eq:R2y1y2}
    R_{2}^{\alpha\beta}(P_{\rm long},P_{\rm out},P_{\rm side})=\frac{C_{2}^{\alpha\beta}(P_{\rm long},P_{\rm out},P_{\rm side})}{[\rho_1^{\alpha}\rho_1^{\beta}](P_{\rm long},P_{\rm out},P_{\rm side})},
\end{align}
because these  are approximately robust against particle (efficiency) losses. 

\section{Acceptance Averaging of the Balance Function}
\label{sec:acceptance}

At RHIC and LHC, the  systems produced in A--A collisions  feature large longitudinal and transverse pressure gradients, it is then  of interest to  carry measurements 
 as a function of  differences $\Delta y= y_1 - y_2$ and   $\Delta\varphi=\varphi_1 -\varphi_2$ simultaneously. The realization of  such measurements in individual particle coordinates  requires the handling of  four dimensional (4D) histograms. Even when using a relatively small number of bins along each dimension, one ends up, computationally, with very large objects that may challenge the capacity of computing nodes used for the data analysis. One additionally also faces a statistical accuracy challenge: the measured pairs are spread across a vast 
number of bins and it may become difficult  to achieve sufficient statistical accuracy across the entire phase space. It is then often desirable to ab initio reduce the 
dimensionality of the measurement by projecting this 4D space onto a 2D space 
$\Delta y$ vs. $\Delta \varphi$. One must then consider how   such projections 
impact the balance functions $B$ and their integrals $I^{\alpha|\bar\beta}$
in measurements featuring a limited acceptance $-y_0 \le y < y_0$.  

\begin{figure}[!ht]
	\centering
	\includegraphics[width=0.5\linewidth,trim={4mm 4mm 4mm 4mm},clip]
	{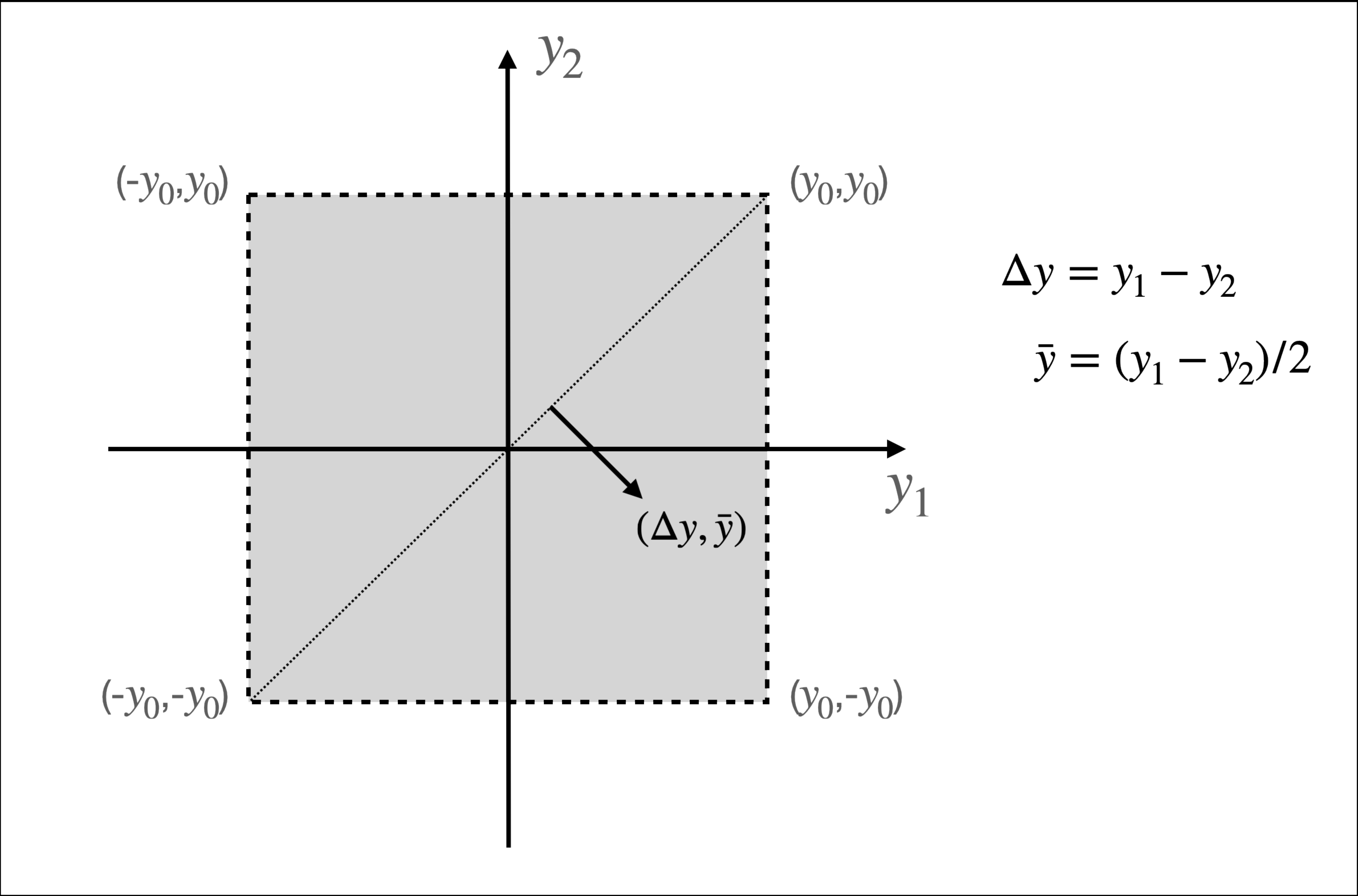}
\caption{Definition of the pair acceptance used in the definition of bound balance functions. }
	\label{fig:Acceptance} 
\end{figure}

In order to carry out computations  in $\Delta y$ and $\Delta \varphi$ coordinates, one first considers the transformations 
\begin{align}
    y_1, y_2 &\rightarrow \Delta y \equiv y_1 - y_2,  \hspace{0.1in} \bar y \equiv (y_1 + y_2)/2, \\
    \varphi_1, \varphi_2 &\rightarrow \Delta \varphi \equiv \varphi_1 - \varphi_2, \hspace{0.1in}\bar \varphi \equiv (\varphi_1 + \varphi_2)/2,
\end{align}
which both feature a Jacobian $J=1$. Densities $\rho_2^{\alpha\beta}(y_{1},y_{2}, \varphi_1, \varphi_2)$ thence transform to  $\rho_2^{\alpha\beta}(\Delta y, \bar y, \Delta\varphi, \bar\varphi)$ 
according to 
\begin{align}
\label{eq:rho2trans}
    \rho_2^{\alpha\beta}(\Delta y,\bar y,\Delta\varphi, \bar\varphi) &= \int dy_1 \int dy_2 
      \hspace{0.05in}\rho_2^{\alpha\beta}(y_{1},y_{2}, \varphi_1, \varphi_2) \hspace{0.05in}\delta(\Delta y-y_1+y_2)  \delta(\bar y-(y_1+y_2)/2.0) \\ \nonumber
      &\times \delta(\Delta \varphi-\varphi_1+\varphi_2)  \delta(\bar \varphi-(\varphi_1+\varphi_2)/2.0). 
\end{align}
Measurements of  $B^{\alpha|\beta}(\Delta y, \Delta\varphi)$ can be carried out as simple projections of the 4D space spanned by $y_1$, $\varphi_1$, $y_2$, $\varphi_2$ or averages across the acceptances $\bar y = \left(y_1 + y_2 \right)/2$ and $\bar \varphi = \left(\varphi_1 + \varphi_2 \right)/2$. Obtaining simple projections is trivial given it suffices to fill histograms of the two densities in terms of the $\Delta y$ and $\Delta\varphi$ coordinates, e.g.,  
\begin{equation}
    \rho_{2}^{\alpha\beta}(\Delta y) \equiv   \int_{\Omega} d\bar y \hspace{0.05in} \rho_{2}^{\alpha\beta}(\Delta y,\bar y),
\end{equation}
However,  such projections emphasize small values of $\Delta y$, e.g., $\Delta y\approx 0$, of the two-particle phase space at the expense of regions with $\Delta y \approx 2 y_0$  near the edge of the acceptance. It is thus advantageous to consider averages across the $\bar y$ acceptance 
as follows
\begin{equation}
    \bar\rho_{2}^{\alpha\beta}(\Delta y) \equiv \frac{1}{\Omega(\Delta y)} \int_{\Omega} d\bar y \hspace{0.05in} \rho_{2}^{\alpha\beta}(\Delta y,\bar y) = \frac{1}{\Omega(\Delta y)} \rho_{2}^{\alpha\beta}(\Delta y),
\end{equation}
where the over-bar in  $\bar \rho$ represents the averaging across  $\bar y$ and $\Omega(\Delta y)$ is the width of the acceptance in $\bar y$ at the given $\Delta y$. For a square and symmetric two-particle acceptance, $-y_0 \le y_1, y_2 < y_0$,
as illustrated in Fig.~\ref{fig:Acceptance}, the value of  $\Omega(\Delta y)$ amounts to
\begin{equation}
    \Omega(\Delta y) = 2 y_0 - |\Delta y|.
\end{equation}
The function  $\Omega(\Delta y)$ is often called acceptance factor. It should be clear, however, that its use does not constitute an acceptance ``correction" but involves acceptance averaging along $\bar y$.

Projections of balance functions $B^{\alpha|\bar\beta}(y_1,y_2)$  onto $\Delta y$ are carried in the same way,
and one distinguishes straight  and acceptance averaged projections 
denoted 
\begin{align}
    B^{\alpha|\bar\beta}(\Delta y) &\equiv   \int_{\Omega} d\bar y \hspace{0.05in} B^{\alpha|\bar\beta}(\Delta y,\bar y), \\
    \bar{B}^{\alpha|\bar\beta}(\Delta y) &\equiv \frac{1}{\Omega(\Delta y)} B^{\alpha|\bar\beta}(\Delta y),
\end{align}
respectively, with similarly formed expressions for $B^{\bar\alpha|\beta}(\Delta y)$ and $\bar B^{\bar\alpha|\beta}(\Delta y_1)$. Evidently, these expressions can be used to compute balance functions based on correlation functions, e.g.,  $C_2^{\bar\alpha|\beta}(\Delta y)$, given by Eqs.~(\ref{eq:B2+-y1y2NoQ-C}, \ref{eq:B2-+y1y2NoQ-C}), or normalized cumulants, represented in Eqs.~(\ref{eq:B2alphaBarBetaVsR2Sym}, \ref{eq:B2BarAlphaBetaVsR2Sym}).
By construction, integrals of $B^{\alpha|\bar\beta}(\Delta y)$, $B^{\bar\alpha|\beta}(\Delta y)$,  yield results identical to those obtained with densities and correlation functions
$C_2^{\bar\alpha|\beta}(y_1,y_2)$. However, integrals of acceptance 
averaged balance functions 
$\bar B^{\alpha|\bar\beta}(\Delta y)$, $\bar B^{\bar\alpha|\beta}(\Delta y)$  do not given they feature the acceptance  factor $\Omega(\Delta y)$ in their definition. Balance function integrals can nonetheless
be recovered by inserting this acceptance factor explicitly in the BF integral as follows
\begin{align}
    I^{\alpha|\bar\beta}(\Omega) = \int_{\Omega} \Omega(\Delta y)  \bar{B}^{\alpha|\bar\beta}(\Delta y) d\Delta y.
\end{align}

\section{Balance functions and the $\nu_{\rm dyn}$ observable}
\label{sec:BalFctVsNudyn}

The $\nu_{\rm dyn}$ observable was initially developed and used for the study of net charge fluctuations~\cite{Pruneau:2002yf}. As such, it corresponds to the ``dynamical" or non statistical  components of net charge fluctuations. It can however also be used for the study of the relative abundance fluctuations of particles species $\alpha$ and $\beta$. In that context, it is most succinctly written as a combination of   normalized  cumulants,  according to 
\begin{equation}
\nu_{\rm dyn}^{\alpha\beta}  = R_2^{\alpha\alpha} + R_2^{\beta\beta} -2 R_2^{\alpha\beta}, 
\end{equation}
with  $R_2^{\alpha\beta}$ correlators defined and  computed according to Eq.~(\ref{eq:normalizedFactorialCumulants}). 
In the context of studies of net charge fluctuations within the  acceptance $\Omega:-y_0 \le y < y_0$, the above reduces to 
\begin{equation}
	\label{eq:nudynPM}
\nu_{\rm dyn}^{+-}(\Omega)  = R_2^{++}(\Omega) + R_2^{--}(\Omega) -2 R_2^{+-}(\Omega),
\end{equation}
with 
\begin{align}
R_2^{++}(\Omega) &= \frac{\langle N_+(N_+-1)\rangle}{\langle N_+\rangle^2} - 1, \\ 
R_2^{--}(\Omega) &= \frac{\langle N_-(N_--1)\rangle}{\langle N_-\rangle^2} - 1, \\
R_2^{+-}(\Omega)  &= \frac{\langle N_+N_-\rangle}{\langle N_+\rangle\langle N_-\rangle} - 1,
\end{align}
in which $\langle N_+(N_+-1)\rangle$,  $\langle N_-(N_--1)\rangle$, and 
$\langle N_+N_-\rangle$ correspond, respectively, to average number of positive  particle  pairs $++$, average number of negative particle pairs $--$, and average number of unlike sign pairs $+-$ detected with the acceptance $\Omega:-y_0 \le y < y_0$. 

We next verify that the above expression for $\nu_{\rm dyn}^{+-}(\Omega)$
is approximately equal to charge  BFs computed with Eqs.~(\ref{eq:B2alphaBarBetaIntNoQ}, \ref{eq:B2BarAlphaBetaIntNoQ}).  To this end, we write BF integrals $I(\Omega)$ according to 
\begin{align}
	I(\Omega) &= \frac{1}{2} \left[ F_1^+ R_2^{+-} + F_1^- R_2^{-+}  - F_1^- R_2^{--}  - F_1^+ R_2^{++} \right]. 
\end{align}
Defining $\omega(\Omega) = F_1^-/F_1^+$, and acknowledging that $R_2^{+-}  = R_2^{-+}$, we divide the above expression by $-F_1^+/2$ and get
\begin{align}
		\label{eq:B2sVsNudynExact}
	-\frac{2I^s}{ F_1^+} &=  \left[    \omega   R_2^{--}  + R_2^{++} -\left( 1 + \omega \right) R_2^{+-}\right],
\end{align}
where we have omitted the dependence on $\Omega$ to simplify the notation. This expression 
 reduces to    $-\nu_{\rm dyn}^{+-}$, given by Eq.~(\ref{eq:nudynPM}), in the limit $\omega(\Omega)\rightarrow 1$ approximately valid at high collision energy for light particles.  Denoting the total average charged particle multiplicity $\langle N\rangle \equiv F_1^+ +  F_1^- =  \langle N_1^+\rangle + \langle N_1^-\rangle$, one thus  recovers the known result
\begin{align}
	\label{eq:B2sVsNudyn}
		I^s &= - \frac{\langle N\rangle }{4}  \nu_{\rm dyn}^{+-}
\end{align}
valid in that limit~\cite{Pruneau:2002yf}. It is important to note that at SPS and RHIC energies, or  even at LHC energy, the limit $\langle N_1^+\rangle = \langle N_1^-\rangle$ is not perfectly achieved. The precision of  the approximation, Eq.~(\ref{eq:B2sVsNudyn}), predicated on $\omega(\Omega)\rightarrow 1$, must thus be
explicitly verified, relative to the correct expression, Eq.~(\ref{eq:B2sVsNudynExact}).

\section{Summary}
\label{sec:summary}

We examined the nominal definition of general charge balance function~\cite{Bass:2000az,Jeon:2001ue} and found that it is advantageous to define two complementary balance functions based on differences of conditional densities of like-sign and unlike-sign pairs of particles. 
We first proceeded to define integral balance functions and showed that in order to account for a system's charge, the balance functions must include a term equal to the difference of positively and negatively charged particle multiplicities. We next showed that differential balance functions $B^{+|-}$ and $B^{-|+}$ defined from differences of conditional densities can also properly account for the system's net charge provided one adds   the difference of positive and negative densities to their definitions. We further showed that such charge balance functions can be generalized to any combinations of species $\alpha$ and $\beta$. We showed, in particular, that such general balance functions also account for finite net charge of the collision system being considered 
provided they include the density  difference  $\rho_1^{\alpha}(y) - \rho_1^{\bar\alpha}(y)$.
We derived the simple sum rules, Eq.~(\ref{eq:sumRuleGeneric+}, \ref{eq:sumRuleGeneric-}, \ref{eq:SumRuleAlphaBeta}) that show that the sum of BFs of particle pairs $\alpha|\bar\beta$ feature an integral across the full phase space that converges to unity. 

Additionally, we also showed charge BFs can be straightforwardly extended to baryon, strangeness, and charm BFs provided one accounts for the baryon,  strangeness, and charm density rather than the particle density. As such, general balance functions could provide a path to a better and deeper understanding of the evolution of systems formed in pp, p--A, and A--A collisions. Moreover,  although not explicitly discussed in this work, it is clear that measurements of balance functions within jets could potentially also yield a better understanding of the structure of jets and their modification in A--A collisions relative to those observed in pp collisions.

Finally, we derived expressions for  bounded balance functions, i.e., balance functions measured in a specific acceptance, based on either densities $\rho_2^{\alpha\beta}$ or normalized correlation functions $R_2^{\alpha\beta}$. We showed that balance functions based on difference variables $\Delta y$ and $\Delta \varphi$ may be computed as straight projections from 4D space $\{y_1,\varphi_1, y_2, \varphi_2\}$ or as weighted averages across the pair rapidity average $\bar y=(y_1+y_2)/2$. We also derived a general formula that connects the integral of charge balance functions and the $\nu_{\rm dyn}^{+-}$ observable.

We have shown that general BFs $B^{+|-}$ and $B^{-|+}$ must include 
the density  difference  $\rho_1^{\alpha}(y) - \rho_1^{\bar\alpha}(y)$ to yield integrals that properly account for the net charge of the collision system considered. But given ratios of particle and anti-particle yields tend towards unity in the central rapidity region, at top RHIC energy and at LHC, one may wonder, however,  whether the inclusion of this term is absolutely essential and whether measurements based on the nominal conditional density difference would constitute reasonable approximations of the correct results. We have also shown that measurements of general balance functions based on $R_2^{\alpha\beta}$ may be carried out based on various experimentally driven approximations. The impact of the omission of the density difference and  $R_2^{\alpha\beta}$ based approximations shall be explored in detail in future works.

\newenvironment{acknowledgement}{\relax}{\relax}
\begin{acknowledgement}
	\section*{Acknowledgements}
	The authors thank Drs. Igor Altsybeev, Peter Christianssen, Scott Pratt, and Sergei Voloshin  for insightful discussions and their suggestions. SB acknowledges the support of the Swedish Research Council (VR) and the Knut and Alice Wallenberg Foundation. This work  was also supported in part by the United States Department of Energy, Office of Nuclear Physics (DOE NP), United States of America, under grant No.  DE-FG02-92ER40713. 
\end{acknowledgement}

\bibliography{main}

\end{document}